\documentstyle[pre,preprint,aps,epsf]{revtex}

\draft
\tighten

\begin{document}

\author{Yu.E.Kuzovlev \\
Donetsk Physics-Technology Institute of NASU\\
ul. R.Luxemburg 72, 340114 Donetsk, Ukraine\\
e-mail: kuzovlev@kinetic.ac.donetsk.ua}

\title{Penetration and transformations of vortices in bulk
current-carrying superconductors}
\maketitle

\begin{abstract}
The equations of viscous evolution of 3D arbitrarily shaped vortices in
an isotropic type II superconductor and necessary boundary conditions
are formulated in the frame of London approximation. The theory is
applied to analyse characteristic scenaria of vortex penetration into
current-carrying thick plate or bulk samples with another geometry.
It is shown that regarding of the surface transport current value a
vortex penetrates either as "flexible stick" or as "elastic thread".
The latter regime is accompanied by giant stretching of the
vortex core along the current-induced surface magnetic field. This
geometrical transformation leads to decrease of viscous friction and
large increase of the vortex drift velocity as compared with the
stick-like regime. As a result, the vortex first winds round the sample
cross-section and forms a ring-like curve, and only later begins to
move deep into the sample interior. The analytical estimates of
the vortex shape and stretching and its velocity are obtained.
\end{abstract}

\section{INTRODUCTION}

Not long ago a variety of complex magnetic structures formed by many
strongly curved and entangled vortices was discovered in bulk
superconductors [1]. The origin of these structures can not be explained if
treat the motion of vortices like that of stick-like objects.
It is necessary to consider the evolution of three-dimensional magnetic
flux lines with potentially arbitrary shape.

For the first, it is useful to investigate the motion and shaping of a
single vortex but interacting with a surface supercurrent which represents
either transport current or Meissner current induced by external field. As
far as we know, even this simple task previously was not under careful
consideration.

Of course, a vortex never lives alone, without interactions with other
vortices, and no stable many-vortex structure could exist without mutual
repulsion of vortices. However, one can suppose that the scenario of
magnetic flux penetration into a current-carrying bulk superconductor should
be dominated only by vortex interaction with the surface current distribution,
i.e. eventually by geometry of superconducting sample, not by the
vortex-vortex interaction. The latter can not seriously affect this scenario,
merely because it itself is unable to ensure a deep penetration at all.

To prove this statement, let us imagine the steady flow of vortices which
arise at the flat current-carrying boundary of half-infinite superconductor
and then move deep into the sample due to their repulsion. Clearly, because
of viscous character of vortex motion, such the flow needs in nonzero
gradient of concentration of vortices. As a consequence, both the
concentration and the local drift velocity of vortices must be decreasing
functions of the depth. Hence, their product is not constant, that is the
magnetic flux conservation can not be satisfied. This discrepancy means that
no steady flow could be supported by the inter-vortex forces only. In
particular, it is impossible to realize the stationary lasting penetration
of vortices from infinite flat boundary parallel to external magnetic field.

Therefore, the only force what can push a vortex through the sample interior
is nothing but self-action of vortex caused by its distorsion. But in order
to involve this force into the evolution, the vortex must feel the shape of
the sample boundary.

Hence, the true picture looks as follows.
After nucleation in a surface layer with thickness of order of
London penetration depth $\lambda $ , a vortex firstly expands over the
sample boundary remaining in this layer. At this stage only the end
fragments of vortex are factually moving. The ends slide along the boundary,
and the resulting shape of vortex core reflects that of boundary. This process
lasts until the curvature of main middle part of the core becomes
sufficiently strong in order to cause the deepening of vortex as a whole.

In view of these reasonings, the geometry of steady transport of magnetic
flux into a bulk supeconductor looks rather insensible to inter-vortex
interactions and thus can be testified in terms of a single vortex, at
least if not consider details of vortex nucleation and processes like
annihilation and reconnections of vortices [1] which take place deep inside
the sample.

For example, many aspects of resistivity in supercurrent-carrying wires can
be described as evolution of ring-like vortices as if thats instantly arise
near the boundary, then contract independently one on another and finally
self-annihilate [2]. However, more correct consideration should include the
first stage when vortex transforms from small nucleus into a ring. We shall
see that in fact this stage may result also in a non-ring penetration geometry,
and more detailed theory can predict what the scenario realizes
under given transport current value and sample dimensions.

Though a lot of works were published previously touching upon a role of
vortex distorsions, for instance, under a pinning by randomly distributed
centers [3], always some preliminary restrictions of the vortex
geometry were attracted. In the present work the general equations of
evolution of arbitrarily curved vortex lines in isotropic superconductors
are formulated and analysed. We shall especially discuss the true
formulation of boundary conditions for these equations.

It will be shown that in a sample whose dimensions noticably exceed
$\lambda $ the vortex can penetrate either as flexible stick or as elastic
(similarly to a rubber thread attracted by its ends through water).
of surface supercurrent. The latter case occurs only if surface
supercurrent exceeds $H_{c1}c/4\pi $  (in CGS units) and
is characterized by giant stretching of the vortex core along the
sample boundary in the direction parallel to drift of the ends.
The stretching is accompanied by decrease of both the vortex
energy and viscous dissipation per unit drift velocity, and
results in strong increase of the vortex drift velocity under given
transport current. In the framework of this scenario, the vortex core
firstly tranforms into a ring-like curve winding round the wire
cross-section (or into a spiral, if there is an external magnetic
field parallel to current), and only later the vortex begins to cut
the wire and enter deep into its interior.
This general picture is in agreement with the known simplific model
of magnetic flux penetration into round wires. Additionally,
our approach allows to scope very different stages of vortex evolution in
unified manner and obtain quantitative estimates for each stage.

\section{LONDON APPROXIMATION}

We shall confine ourselves by the London approximation. Of course, it would
have no sense if one could not apply it to actually moving vortices. But in
any case the requirement must be satisfied that characteristic velocity $u_0$
of viscous vortex motion influenced by magnetic fields comparable with
low critical magnetic field $H_{c1}$ , must be significantly smaller than
the speed of electromagnetic waves.

The velocity scale $u_0$ can be naturally estimated as
\[
u_0\equiv \mu \varepsilon /\lambda
\]
where
\[
\varepsilon =\Phi _0H_{c1}/4\pi
\]
is the self-energy per unit length of long straight-line vortex,
$\varepsilon /\lambda $ is the characteristic scale of Lorentz force also
related to unit length, and $\mu $ is mobility of the vortex core.
Below it will be seen that so defined $u_0$ really serves as the velocity
unit. If combine this definition and the known relations [4]
\[
\frac{c^2}{\Phi _0\mu }\sim \sigma _nH_{c2}\,\,\,,\,\,\,\sigma _n\sim
\frac \hbar \Delta (\frac c\lambda )^2
\]
with standard notations, $\sigma _n$ being the normal conductivity and
$\Delta \sim 2k_BT_c$ being the order parameter, one obtains
\[
u_0\sim \frac{\lambda k_BT_c}{2\pi \hbar }\frac{H_{c1}}{H_{c2}}
\]

As a typical example, at
$T_c\sim 100\,K$ , $\lambda \sim 3\cdot 10^{-5}\,cm$ and
$H_{{c2}}/H_{c1}\sim 500$ , one gets the estimate $u_0\sim 10^5\,cm/s$ .
This value looks small enough to allow for applicability
of quazi-static London approximation.
In fact, such an approach was used in large number of works on motion of
separate vortices as well as vortex lattices. The obvious exception is very
dense lattice, with small inter-vortex distancies of order of coherence
length. But our present subjects of interest are far from such complications.

\section{EVOLUTION EQUATIONS}

In the framework of London approximation, the free energy $E$ of vortex,
placed into a given surroundings, is completely determined by the shape of
its core, $R(p)=\{X(p),Y(p),Z(p)\}$ , with $X,Y,Z$ being coordinates of
the core points and $p$ being a scalar parameter. In accordance with the
principles of mechanics and nonequilibrium thermodynamics, the simplest
equation of a massless viscous evolution of the core line looks as
\begin{equation}
\mu^{-1} \partial R/\partial t= f(R)
\end{equation}
with Lorentz force on the right-hand side and friction force on the
left, both being related to unit core length.

By its sense, the parameter $\mu ^{-1}$ is the effective drag coefficient
which is determined by all the dissipative energy losses conjugated with the
core motion. Generally, there are at least two sorts of dissipative
processes accompanying the motion (see, for example, the review [4]),
namely, relaxation of the order parameter and normal currents
induced by time-dependent own magnetic field of the vortex. A concrete
expression for $\mu $ can be derived from more detailed theory, for
instance, from the Ginzburg-Landau functional approach, under its reduction
to London approximation [4]. The reduction is possible because normal
self-current of moving vortex and corresponding dissipation are located
mainly in a close vicinity of the core line, at distance comparable with
coherence length. After the transition to London's description, the effect
of normal currents becomes hidden in $\mu $ , but these currents give no
contribution to the Lorentz force [4]. Therefore, the reduction results in
the identity whose meaning is balance of friction force and Lorentz
force, as it is stated by the Eq.1, with $f(R)$ being determined only
by supercurrents.

To write $f(R)$ , one has not to evaluate the supercurrent
distribution. Instead, as in general in mechanics and statistical
thermodynamics, $f(R)$ can be expressed by means of $E$'s variation under a
small displacement of a local core fragment, that is as the functional
derivative $\delta E/\delta R(p)$ . However, the latter itself is not
invariant with respect to arbitrary (non-degenerated) transformations of the
parametrization $R(p)$ and to physical dimensionality of $p$ . In case of
isotropic media, the only true invariant expression for the Lorentz force is
\begin{equation}
f=-\frac{dp}{dL}\frac{\delta E}{\delta R(p)}=-\left|
\frac{\partial R}{\partial p}\right| ^{-1}
\frac{\delta E}{\delta R(p)}=\frac{\Phi _0}c\left[ J\times N\right]
\end{equation}
Here the vector $\partial R/\partial p\equiv R^{\prime }$ is locally
parallel to the core, $dL=|R^{\prime }|\,dp$ is the differential of the core
length,
$N\equiv R^{\prime }\left| R^{\prime }\right| ^{-1}=\partial R/\partial L$ ,
and $J$ is the density of full effective supercurrent which streams around
core and pushes a given core fragment.

The energy $E=E\{R(p)\}$ includes self-interaction of vortex and
its interaction with surroundings, in particular, with other vortices.
Correspondingly, in general $J$ consists of external currents and
self-current of vortex determined by its distorsion.
The Eqs.1 and 2 could be directly extended to a number of interacting
vortices. Besides, in principle, one may add
into $E$ also interactions with pinning potentials. However, below we
are interested only in motion of separate vortex in absence of pinning.

The parameter $p$ enumerates strictly the core points. But in practice it is
preferable to use another kind of parametrization, concretely, to introduce
the parameter $q$ which enumerates some suitable continuum of surfaces
$Q(r)=q$ , $r=\{x,y,z\}$ , each possessing only one intersection with
core line. The connection between $p$ and new parameter $q$ is
implied by the obvious relation $Q(R(p(q,t),t))=q$ , and simple algebraic
manipulations lead to the modified form of the evolution equations,
\begin{equation}
\frac{\partial R}{\partial t}=\mu
\left[ 1-\frac{\partial R}{\partial q} \otimes
\frac{\partial Q(R)}{\partial R}\right]f(R)\,\,,\,\,\,
\,f(R)=-\left| \frac{\partial R}{\partial q}
\right| ^{-1}\frac{\delta E}{\delta R(q)}
\end{equation}
where the symbol $\otimes $ denotes the tensor product of two vectors.

These equations describe how the intersection points marked by $q$  move
along the corresponding surfaces $Q(r)=q$ . Clearly, this is factually
two-dimensional motion. This feature becomes quite obvious
if it is possible
to identify $q$ as one of cartezian coordinates, that is to use parallel
planes as the marking surfaces. For instance, if thats are XY-planes, $q=Z$
and $Q(r)=z$ , then the Eqs.3 reduces to the equation
\begin{equation}
\frac \partial {dt}\left(
\begin{array}{c}
X \\
Y
\end{array}
\right) =-\frac \mu {\sqrt{1+X^{\prime 2}+Y^{\prime 2}}}\left(
\begin{array}{cc}
1+X^{\prime 2} & X^{\prime }Y^{\prime } \\
Y^{\prime }X^{\prime } & 1+Y^{\prime 2}
\end{array}
\right) \left(
\begin{array}{c}
\delta E/\delta X(Z) \\
\delta E/\delta Y(Z)
\end{array}
\right)
\end{equation}
with shortened notations $X^{\prime }\equiv \partial X/\partial Z$ ,
$Y^{\prime }\equiv \partial Y/\partial Z$ . The Eq.4 describes the time
evolution of $X$ and $Y$ coordinates of the core points marked with
their $Z$-coordinate.

\section{BOUNDARY CONDITIONS AND LOCAL APPROXIMATION}

In absence of pinning and more vortices, the vortex energy $E=E_s+E_i$
consists of two parts: the energy $E_i$ of the vortex interaction with
transport or Meissner supercurent and the self-energy $E_s$. Therefore, the
current in the Eq.2 also can be devided into two parts, $J=J_s+J_i$ .

Formally, $E_s$ is a complicated spatially non-local functional [5]
depending on both the core configuration $R(p)$ and the shape of sample.
Among other factors, $E_s$ includes the vortex interaction with the sample
boundary what can be interpreted as attraction of the end fragments of the
core to their mirror images placed outside superconductor.

But, if the curvature radius of the core everywhere is not too small as
compared with $\lambda $ , and besides, if the core nowhere is too close to
itself, then the so-called local approximation is possible,
\begin{equation}
E_s\approx \varepsilon L=\varepsilon \int \left| dR(p)\right|
\end{equation}
where $L$ is the core length. This well known approximation was argued and
used as long ago as in 1968 by Galaiko [6], and later by many other authors
(in particular, in [2-5]). Our own computer simulations showed that the
relative error of evaluation of self-action force by means of local
approximation does not exceed a few percents even if the curvature radius is
as small as $0.1\lambda $ .

In the local approximation the Eqs.1 and 2 take the form
\begin{equation}
\frac{\partial R}{\partial t}=\mu \frac{\Phi _0}c\left[ (J_s+J_i)\times
N\right] \,\,\,\,\,,\,\,\,\,\,J_s=\frac{cH_{c1}}{4\pi }\left[
N\times \frac{\partial ^2R}{\partial p^2}\right] \left|
\frac{\partial R}{ \partial p}\right| ^{-2}
\end{equation}
Here $J_s$ is the self-current what flows through the very core. Its
absolute value is inversely proportional to the local curvature
radius of core.

However, the local approximation needs to be accompanied by correct boundary
conditions. The true conditions should take into account the vortex
interaction with superconductor boundary. There are two ways to show that
this interaction results in the orthogonality of the end fragments of core
to the boundary. Thought these conditions are known at least since [6],
sometimes thats are neglected, so it is desirable to present more
argumentation.

First, let us note that the force vector $f(R)$ is always perpendicular to
the local core direction. Indeed, any variarion $\delta R$  parallel to this
core direction, $\delta R\parallel \partial R/\partial p$ ,
merely is identical to a change of parametrization,
without factual change of the shape, so
it has no physical meaning and should result
in $\delta E=0$ (therefore the
last expression in (2) always is consistent with previous ones). The same is
seen from (6). As a consequense, any core point displaces perpendicularly to
the core, in particular, the end points do which are placed
just on the boundary.
Hence, we must conclude that the end fragments
always are oriented to be orthogonal with respect to the boundary.

Secondly, the non-orthogonality would mean that the contour formed by
core and its mirror image is broken at the end point, i.e. has infinitely
small curvature radius here. From the point of view of exact $E_s$ [5],
if such a sharp "knee" occured it
would cause infinitely strong Lorentz self-action force and
consequently would be immediately straightened thus restoring the
orthogonality.

But, we must to underline that the principal conclusions to be deduced do
not refer to the local approximation and can be derived from general
non-local Eqs.1 and 2 only.

\section{STICK-ELASTIC VORTEX TRANSFORMATION\\ IN CURRENT-CARRYING PLATE}

To avoid a complicated mathematics, we confine ourselves by
simplific superconductor geometry. Consider the vortex evolution in an
infinitely wide plate, $-D<Z<D$ , without pinning but in presence of
transport surface supercurrent uniformly distributed over the boundary
planes and obeying the London equation. If this current flows along Y-axis
then
\[
J_x=J_z=0\,\,\,\,\,,\,\,\,\,J_y=\frac c{4\pi \lambda
}H_{c1}j(Z)\,\,\,\,,\,\,\,j(Z)\equiv h\frac{\cosh (Z/\lambda )}{\cosh
(D/\lambda )}
\]
with $h$ being the dimensionless measure of current density.

Let initially the vortex pierces the plate in Z-direction being described
with $R=\{X(Z,0)=0,0,Z\}$ . It has similar orientation soon after nucleation
near the edge of a real finite plate. Then, due to obvious spatial symmetry,
the vortex will remain inside the XZ-plane $Y=0$ and keep only one
intersection with any of XY-planes. In this situation the Eq.4 can be
applied and, besides, reduced to only equation for
X-coordinate, $X=X(Z,t)$ , as a function of time and Z -coordinate:
\begin{equation}
\frac{\partial X}{\partial t}=-\mu
\sqrt{1+X^{\prime 2}}\frac{\delta E}{\delta X(Z)}
\end{equation}
The energy can be expressed as
\begin{equation}
E=E_s+E_i=E_s-\frac \varepsilon \lambda \int X(Z,t)j(Z)dZ
\end{equation}
where the integral represents
the energy $E_i$ of vortex interaction with transport
current (this expression differs only by some constant from the general $E_i$
representation [5]). In the local approximation (5), the Eq.7 looks as
strongly nonlinear diffusion-type equation
\begin{equation}
\frac{\partial X}{\partial t}=u_0\left[ \lambda
\frac{X^{\prime \prime }}{1+X^{\prime 2}}+\sqrt{1+X^{\prime 2}}j(Z)\right]
\end{equation}
with notation $X^{\prime \prime }\equiv \partial ^2X/\partial Z^2$ and
characteristic velocity $u_0$ introduced in Sec.2.

Here the left side is responsible for the friction, and two terms on the
right-hand side represent the self-action force and transport
current-induced Lorenz force, respectively. Clearly, because of the latter
force both the vortex ends will forwardly move in one and the same direction
(to opposite edge of the plate), while the middle of vortex will be more or
less backward, and the larger is transport current the longer should be the
distance $\Delta X = X(\pm D,t)-X(0,t) $ (below termed vortex stretching).

Some predictions of further vortex behaviour can be deduced merely from the
energy expression (8). Just after start the middle is still in rest. As the
Eq.8 shows, in thick plate ($D>>\lambda $) the unit displacement of every
end leads to the $E_i$ 's decrease by $h\epsilon $ . At the same time, the
corresponding lengthening of each of two symmetrical core branches results
in the $E_s$ 's increase by $\epsilon $ per unit length. Consequently, if
$h>1$ then the total energy decreases,
and the vortex stretchening along the
drift direction becomes profitable. The lengthening process should last
until the curvature of the most backward central part of the core becomes so
large that the self-action force makes this part moving as quickly as the
ends do.

Thus, at $h>1$ the vortex gets over the friction like elastic in water.
Oppositely, at $h<1$ the stretchening is energetically unprofitable, and the
vortex should move as deformed flexible stick. The transition from this
stick-like behaviour to elastic-like one, when
transport current increases from $h<1$ to $h>1$ ,
is the example of so-called "nonequilibrium phase transitions".

Let us consider the steady drift of vortex as a whole, without change of
shape. The corresponding solution on the Eqs. 7 or 9 can be written as
$X(Z,t)=ut+X(Z)$ . The stationary shape $X(Z)$ and the drift velocity
$u=u(h,D)$ should be obtained from (7) or (9) with the help of above
discussed orthogonality boundary conditions
\[
\frac{dX}{dZ}(\pm D)=0
\]
Besides, due to the mirror symmetry,
the condition $X^{\prime }(\pm 0)=0$ should be satisfied.

In this steady nonequilibrium state the self-energy $E_s$ is constant,
therefore, the work $M_j$ produced by transport current per unit time,
\[
M_j=-\frac{dE_i}{dt}=\frac{u\Phi _0}c\int_{-D}^DJ_y(Z)dZ=2uh\varepsilon
\tanh (D/\lambda )
\]
coinsides with the energy dissipation per unit time $M_d$ . In accordance
with (1) and (2),
\[
W_d=\frac 1\mu \int \left| \frac{\partial R}{\partial t}\right| ^2dL=\frac
1\mu \int \left( \frac uQ\right) ^2QdZ=2\delta u^2/\mu
\]
where the notations
\[
Q\equiv \sqrt{1+X^{\prime 2}}=\frac{dL}{dZ}\,\,\,,\,\,\delta \equiv
\int\limits_0^D\frac{dZ}Q\,
\]
are introduced. We took into account that actual displacement of the core
always is locally perpendicular to its orientation. Only such displacements
are physically meaningful and really cause the friction. Therefore, the
drift velocity and the local core velocity are connected by the relation
\[
\left| \frac{\partial R}{\partial t}\right| =\frac uQ
\]
Evidently, the factor $Q$ determines at one and the same time local
orientation of the core and degree of its stretching.

Hence, the equality $M_d=M_j$ yields
\begin{equation}
U\equiv \frac u{u_0}=\frac{h\lambda }\delta
\tanh (D/\lambda )\approx \frac{ h\lambda }\delta
\end{equation}

In view of above reasonings, at $h<1$ the vortex stretchening is weak,
therefore, $X^{\prime 2}$ is comparable with unit, $Q\sim 1$,
$\Delta X\sim D$ and $\delta \sim D$.
Then the Eq.10 shows that in this stick-like regime
$U\sim h\lambda /\delta \approx h\lambda /D<<h$ ,
i.e. the drift velocity is
inversely proportional to the plate thickness. This is quite natural,
because the surface current-induced Lorentz force acts only on the ends,
while the friction almost equally acts on any core fragment.

In general, the parameter $\delta $ serves as the effective plate
half-thickness. Obviously, always $\delta <D$ . In the stretched
elastic-like regime in thick plate anywhere at $D-|Z|>> \lambda $ the
inequalities $\left| X^{\prime }\right| >>1$ and $Q>>1$ take place.
Hence, $\delta <<D$ and what is essential it becomes almost
insensitive to thickness. As a consequence, both the drift velocity and
mobility $u/h$ strongly increase as compared with stick-like regime and both
become independent on thickness (below we shall see that $\delta \sim
\lambda $ and $U\sim h$ , i.e. $U$ becomes approximately $D/\lambda $ times
larger). According to the $M_d$'s expression, the matter is that thought
the energy dissipation $Q$ times increases due to the core lengthening this
effect is overpowered by its $Q^2$ decrease because of $Q$ times decrease of
the factual core velocity $\left| \frac{\partial R}{\partial t}\right| $ .
As the result, the vortex stretching leads to smaller friction and smaller
entropy production, under fixed vortex velocity, and to larger velocity
ander fixed transport current. The picture looks as if most part of core
slides along itself, but this process does not mean a real motion of core
and so does not cause a friction and dissipation.

\section{DRIFT OF THE VORTEX ENDS}

To be convinced in what was said, let us consider vortex shape in the
stretched elastic-like regime. Because at $\Delta X>>D$ most part of the
core inevitably has a small curvature, it can be considered with neglecting
the self-action. Then any of the Eqs.7 and 9 reduces to
\begin{equation}
U\approx \sqrt{1+X^{\prime 2}}j(Z)
\end{equation}
Here from the characteristic exponential asymptotics does follow,
\begin{equation}
X(Z)\approx -\frac{\lambda U}h\left[ \exp \left( \frac{D-Z}\lambda \right)
-1\right]
\end{equation}
Here $Z>0$ , $X(-Z)=X(Z)$ , and for definitness the position $X=0$ is
prescribed to the end fragments. It is easy to verify that corresponding
self-action contribution in the Eq.9 indeed is negligibly small as compared
with the current-induced force.

We can get a rough estimate of the stretching if put on $Z=0$ in (12) and
take into account that $U>h\lambda /D$. Then the Eq.12 yields
\[
\Delta X/\lambda \approx \frac Uh\exp \left( \frac D\lambda \right) >\frac
\lambda D\exp \left( \frac D\lambda \right)
\]
Hence, $\Delta X/\lambda $ possesses exponentially strong dependence on
$D/\lambda $ , and it can be giantly large if $D$ exceeds $\lambda $ by an
order of value or more.

In view of this circumstance, the ratio $\Delta X /W $ with $W$ being the
width of a real finite plate, becomes of principal importance. Clearly, if
$\Delta X >> W $ then the steady drift of the vortex as a whole is
impossible: the ends of vortex will achieve the opposite edge before the
displacement of its backward central part will be comparable with $W$ (all
the more, before the velocity of this part becomes equal to that of the
ends).

Consider the drift of the ends in such a non-stationary situation. Because
the vortex lengthening is profitable, this drift can do independently on the
motion of deepened backward part, as if thickness was infinitely large
($D/\lambda \rightarrow \infty $) .
To estimate the drift velocity, let us
multiply the Eq.7 or 9 (with $\frac{\partial X}{\partial t}\Rightarrow u$ )
by $Q^{-1}$ and integrate over variable  $z=D-Z$  from zero to infinity,
with the condition $X^{\prime }(z\rightarrow \infty )=\infty $ which
evidently corresponds to infinitely far backward center. Then both the Eqs.7
and 9 result in
\[
U=\frac \lambda \delta (h-1)\,\,\,,\,\,\delta =\int_0^\infty \frac{dz}Q
\]

To evaluate this integral, note that in accordance with the orthogonality
boundary conditions the shape of the end fragments of the core is parabolic,
for instance, at upper end $X(Z)=X(D)-(Z-D)^2/2\rho $ , with $\rho $ being
the curvature radius at the end point. It follows from the Eq.9 that
$\lambda /\rho =h-U$  .
In this parabolic region the integration divergers
but becomes cut after transition to exponential asymptotics (12). The
estimate of the integral leads to approximate equation
\begin{equation}
U\approx h(h-1)/\{h-1+\ln [2z_0(h-U)/\lambda ]\}
\end{equation}
where $z_0$ is the depth of the crossover point, $z_0\sim 4\lambda $ . The
Eq.12 helps to estimate the end drift velocity in thick plate. Obviously, it
turns into zero at $h\rightarrow 0$, in agreement with $D\rightarrow \infty $
limit of the estimate for stick-like regime. It can be shown that velocity
of the steady drift of the vortex as a whole is only slightly smaller
differing by a multiplier of order of unit.

\section{GIANT VORTEX STRETCHING}

The exponentially large vortex stretching is the most significant
possibility of vortex evolution in thick plate, as well as in bulk samples
in general. Therefore it would be useful to more correctly justify the above
simplific estimate of $\Delta X $ .

Note that $\Delta X>L/2-D$ . Divide both sides of (9) by $j(Z)$ and
integrate from zero to $D$ . This results in
\[
L/2=B-A\,\,,\,\,B\equiv U\int \frac{dZ}{j(Z)}\,\,,\,\,\,\,A\equiv \int
\arctan (X^{\prime })\left| \frac d{dZ}\frac \lambda {j(Z)}\right| dZ
\]
It is easy to notice that
\[
A<A_0\equiv \frac{\pi \lambda }2\left[ \frac 1{j(0)}-\frac 1{j(D)}\right]
\]
so $\Delta X>B-A_0-D$. The calculation of integral $B$ gives
\[
B=\frac{2U\lambda }h\cosh (D/\lambda )\{\arctan [\exp (D/\lambda )]-\frac
\pi 4\}
\]
Then, after simplifications possible due to $D>>\lambda $, one finally
obtains
\begin{equation}
\Delta X/\lambda >\frac{\pi (U-k)}{4h}\exp (D/\lambda )\,
\end{equation}
with $k<1$ . Because $u$ is monotonously growing function of $h$ , the
coefficient in front of exponent is positive if $h$ exceeds some level
larger than unit, for example, if $h>2$ . Hence, ate least at $h>2$ the
vortex stretching is exponentially strong.

This estimate is obtained in the framework of local approximation. More
correct estimate should give a lesser value, because of self-attraction of
the core in middle part of the plate where two symmetrical exponential tails
described by (12) meet one another and form an arc. Such a non-local effect
is most essential just under the specific plate geometry. However, the
non-local correction can not change the shape of the front vortex part where
the non-local interaction is weak as compared with other forces. It is
not hard to show that the latter requirement is
satisfied if $|Z|>Z_0$, where $Z_0$ is the solution on
equation
\[
h\exp [-(D-\left| Z\right| )/\lambda ]\approx \sqrt{\lambda /2\pi \left|
Z\right| }\exp (-2\left| Z\right| /\lambda )
\]
If take into account that the more is $h$ the less is $Z_0$, then this
equation yields  $Z_0<D/3$ . Hence, at $|Z|>D/3 $
the mutual attraction of two core branches can be neglected, and the
asymptotics (12) remains valid. This means that the maximally possible
effect of non-locality is the replacing $D$
in the exponent by $\alpha D$ with $\alpha >2/3$ . Consequently, the lower
bound for the stretching with confidence can be estimated as
\[
\Delta X/\lambda >\frac \lambda D\exp (2D/3\lambda )
\]
Thus, even in the worst case the non-local effects
do not abolish the exponential character of stretching.

For example, if $\lambda \approx 3\cdot 10^{-5}\,cm $
and $h$ equals to a few units, then
even at $D\sim 20\lambda <10^{-3}\,cm$ one gets
$\Delta X >m\lambda \exp(2D/3\lambda ) $ , with $m \sim 1 $ ,
i.e. $\Delta X > 1\,cm $ what exceeds a width of any realistic sample.
Thus at first the vortex should form a ring whose shape approximately copies
that of the sample cross-section. During this process the velocity of
backward deepened core part is primarity determined by its distorsion
which is created in the beginning of stretchening and thus has
curvature radius of order of $D$ . Hence, this velocity is of
order of $u_0\lambda /D$ , and at the moment when the ends will meet
one another the displacement of most backward point will be
yet as small as $\sim \lambda W/Dh<<W$ .

\section{DISCUSSION AND RESUME}

It seems clear that both the above conclusions can be extended to bulk
current-carrying superconductors with another geometry, for instance, to
round wires, if treat $2D$ and $W$ as minimal and
maximal diameters of cross-section of
the wire, respectively.

Due to possibility of giant deformation and stretching of vortices, the
thermodynamically nonequilibrium process of vortex penetration
can promote formation of complicated many-vortex dynamical configurations
which seem rather strange and unprofitable from the point of view of
equilibrium thermodynamics. The presence of an external magnetic field
parallel to transport current should lead to formation of
spiral-like configuration instead of ring-like one and thus especially
ensure the entangling of vortices. The Eq.4 enables us to describe
this scenario in details, if choose Z-axis to be directed along the wire.
Besides, the presence of weak pinning should amplifier the stretching
of vortex and additionly complicate its shaping,
because the motion of deepened part of
vortex is characterized by relatively small forces of order of
$\epsilon \lambda /D $ (much smaller than forces $\sim \epsilon $
what act on the end fragments) and so may be easily held
back by pinning centers.

We would like to underline the role of orthogonality boundary conditions.
In the work [3] the equation was under use similar to
our Eq.9, but boundary conditions was formulated in terms of the tension
of core line. One can see from [3] that such conditions make it
impossible to
consider the case of high surface transport current $>H_{c1} $
corresponding to the elastic-like regime.

To resume, we formulated the
invariant equations of viscous motion of arbitrarily shaped 3D vortex
lines, and applied them to careful analysis of the scenario of
vortex penetration into a thick superconducting sample. As it was
argued, the vortex-vortex interaction does not significantly affect
the penetration process. But, of course, a full description of
resistive state leads to more complicated tasks about vortex-vortex
interactions deep inside the sample.

\,\,\,\,

ACKNOWLEDGEMENTS

I would like to thank Dr. M.Indenbom, Dr. Yu.Genenko and Dr. A.Radievskiy
for usefull discussions.

\,\,\,\,

REFERENCES

1. M.V.Indenbom, C.J.van der Beek, V.Berseth, W.Benoit, G.D'Anna,

A.Erb, E.Walker and R.Flukiger, Nature, 1997, Feb.20 .

2. Yu.A.Genenko, Phys.Rev., B 49, 1994, 6950.

3. Chao Tang, Shechao Feng and L.Golubovich, Phys.Rev.Lett.,
472, 1994, 1264.

4. L.P.Gorkov and N.B.Kopnin, Uspekhi fizicheskikh nauk, 116, 1975, 411
(transl. in English in Sov.Phys.-Usp., 1975).

5. Yu.E.Kuzovlev, Physica, C 292, 1997, 117.

6. V.P.Galaiko, Zh.Teor.Eksp.Fiz., 50 , 1966, 1322.

\end{document}